\documentclass[prd,superscriptaddress,twocolumn,showpacs,%
preprintnumbers]{revtex4}
\usepackage{bm}
\usepackage{graphicx}
\newcommand{\beq}{\begin{equation}}
\newcommand{\eeq}{\end{equation}}
\newcommand{\be}{\begin{eqnarray}}
\newcommand{\ee}{\end{eqnarray}}
\begin{document}
\title{Higher twists in polarized DIS and the
size of the constituent quark}
\author{A.~V.~Sidorov}
\affiliation{Bogoliubov Laboratory of Theoretical Physics,
Joint Institute for Nuclear Research, 141980 Dubna, Russia}
\author{C.~Weiss}
\affiliation{Theory Center, Jefferson Lab, Newport News, VA 23606, USA}
\begin{abstract}
The spontaneous breaking of chiral symmetry implies the presence of a
short--distance scale in the QCD vacuum, which phenomenologically
may be associated with the ``size'' of the constituent quark, 
$\rho \approx 0.3 \, \text{fm}$. We discuss the role of this scale
in the matrix elements of the twist--4 and 3 quark--gluon operators
determining the leading power ($1/Q^2$--) corrections to the moments
of the nucleon spin structure functions. We argue that the flavor--nonsinglet 
twist--4 matrix element, $f_2^{u - d}$, has a sizable negative value of 
the order $\rho^{-2}$, due to the presence of sea quarks with virtualities
$\sim \rho^{-2}$ in the proton wave function. The twist--3 matrix element,
$d_2$, is not related to the scale $\rho^{-2}$. Our arguments support the 
results of previous calculations of the matrix elements in the instanton 
vacuum model. We show that this qualitative picture is in agreement 
with the phenomenological higher--twist correction extracted from 
an NLO QCD fit to the world data on $g_1^p$ and $g_1^n$, which include 
recent data from the Jefferson Lab Hall A and COMPASS experiments. 
We comment on the implications of the short--distance scale $\rho$ for 
quark--hadron duality and the $x$--dependence of higher--twist contributions.
\end{abstract}
\preprint{JLAB-THY-06-468}
\maketitle
\section{Introduction}
\label{sec:introduction}
Polarized deep--inelastic scattering (DIS) has extensively been studied 
in fixed--target experiments with electron and muon beams \cite{summary}.
The main goal of such measurements is the extraction 
of the polarized parton densities in the nucleon from the data in the region 
of approximate Bjorken scaling, where the $Q^2$--dependence is governed by 
perturbative QCD (DGLAP evolution). 
However, most of the precise data are at momentum transfers
$Q^2 \sim \text{few GeV}^2$, where non-perturbative effects 
in the $Q^2$ dependence (power corrections) generally cannot be 
neglected. Estimates of these corrections are needed in order to 
include these data in the QCD analysis. The effect of phenomenological
power corrections on the extracted parton densities has been investigated 
in Refs.~\cite{Leader:2002ni,Leader:2004qv}.

With the polarized DIS data becoming more and more precise, the 
study of the power corrections themselves has emerged as an 
interesting subject. The $1/Q^2$--corrections to the 
lowest non-zero moments of the spin structure functions, 
$g_1$ and $g_2$, are governed by matrix elements of QCD operators
of twist 4 and 3, describing non-perturbative correlations of the
quark and gluon fields in the nucleon \cite{Shuryak:1981pi,Ji:1993sv}:
\be
\langle N | \bar\psi \gamma_\alpha \, g \widetilde{G}^{\beta\alpha} 
\psi | N \rangle &=& 2 \, f_2 \; s^\beta ,
\label{f2_def}
\\
\langle N| \bar\psi  \gamma^{\left\{\alpha\right.} 
g \widetilde{G}^{\left.\beta\right\}\gamma} \psi  
| N \rangle \nonumber 
&=& 2 \, d_2 \; ( p^{\left\{\alpha\right.} p^{\left.\beta\right\}} 
s^\gamma - p^\gamma p^{\left\{\beta\right.} s^{\left.\alpha\right\}} ) 
\nonumber \\
&+& \text{traces},
\label{d2_def}
\ee
where $\widetilde G^{\alpha\beta} = (1/2)
\epsilon^{\alpha\beta\gamma\delta} G_{\gamma\delta}$ is the dual
gluon field strength, and $p^\alpha$ and $s^\alpha$ the nucleon 
four--momentum and polarization vector; we follow the conventions
of Ref.~\cite{Ehrnsperger:1993hh} \footnote{Contrary to 
Ref.~\cite{Ehrnsperger:1993hh} and standard convention, 
we define the coefficient of the twist--4 matrix element, 
$f_2$, Eq.~(\ref{f2_def}), such that it has dimension $(\text{mass})^2$, 
\textit{i.e.}, we do not make it dimensionless by extracting in a factor of 
$M_N^2$ (nucleon mass). We shall see below that the value of $f_2$ is 
determined by a physical mass scale unrelated to the nucleon mass.
Extracting a factor $M_N^2$ would obscure this fact, and burden the
subsequent formulas with unnecessary factors.}. By extracting the 
coefficients of the $1/Q^2$ corrections from the data one can thus obtain
information about the structure of the nucleon in QCD.
The SLAC E155X experiment \cite{Anthony:2002hy} and
the Jefferson Lab Hall A experiment \cite{Zheng:2004ce} have
extracted the twist--3 matrix elements from combined
measurements of the structure functions $g_1$ and $g_2$,
and found surprisingly small values, $d_2 \lesssim \, 10^{-2}$. 
Recent analyses have also attempted to extract the twist--4 matrix element, 
$f_2$, from the power corrections to the first moment of 
$g_1$ \cite{Meziani:2004ne,Osipenko:2004xg,Deur:2004ti,Deur:2005jt};
see Ref.~\cite{Chen:2005td} for a review.

The theoretical estimation of the higher--twist matrix elements
(\ref{f2_def}) and (\ref{d2_def}) 
is a challenging problem, requiring a description of the
nucleon in terms of QCD degrees of freedom. The key question is which
non-perturbative scales govern the quark--gluon correlations measured
by the twist--4 and 3 operators.  This question is intimately related
to the role of the ``vacuum structure'' of QCD in determining the
structure of the nucleon.

There is strong evidence for the existence of a short--distance scale
in the QCD vacuum, significantly smaller than the size of the nucleon. 
It is determined by the characteristic size, $\rho$, of the non-perturbative 
field configurations instrumental in the spontaneous breaking of chiral
symmetry. Numerous observations suggest a ``two--scale picture'' 
of hadron structure,
\begin{equation}
\rho \;\; \ll \;\; R ,
\label{hierarchy}
\end{equation}
where $R$ is a typical radius of the nucleon (say, the charge radius).
Phenomenologically, the short--distance scale may be associated with 
the ``size'' of the constituent quark. In fact, the success of effective
models based on constituent quark degrees of freedom could not be explained
without the hierarchy (\ref{hierarchy}). It is natural to ask what the 
existence of this short--distance scale implies for the quark--gluon 
correlations probed by the twist--4 and 3 operators.

A microscopic model incorporating the short--distance scale
associated with chiral symmetry breaking is the instanton vacuum, 
in which the quarks obtain a dynamical mass, $M$, 
by interaction with a ``medium'' of instantons of characteristic 
size $\rho \approx 0.3\; {\rm fm}$, see 
Refs.~\cite{Diakonov:2002fq,Schafer:1996wv} for a review. 
The fundamental assumption of diluteness of the instanton medium 
(smallness of the packing fraction)
implies that, parametrically, $M \ll \rho^{-1}$. Since the range
of the chiral forces binding the constituent quarks is of the
order $M^{-1}$, this hierarchy translates into a ``two--scale picture'' 
of hadron structure, Eq.~(\ref{hierarchy}).
The higher--twist matrix elements (\ref{f2_def}) and (\ref{d2_def})
were calculated in this model in Refs.~\cite{Balla:1997hf,Lee:2001ug}.
It was found that the flavor--nonsinglet twist--4 matrix element
is determined by the inverse instanton size (\textit{i.e.}, the
inverse size of the constituent quark), $f_2^{u - d} \sim \rho^{-2}$,
and has a sizable negative value. The twist--3 matrix element, $d_2$, 
however, is parametrically suppressed, $d_2 \sim (M\rho)^4$,
in agreement with the experimental data (this prediction was later
confirmed by lattice QCD calculations \cite{Gockeler:2000ja}).
An interesting question is whether these results depend specifically
on the assumption of instantons as the dominant gluonic vacuum 
fluctuations, or whether they already follow from the more general 
``two--scale picture'' of hadron structure.

In this paper we further explore the connection between the 
short--distance scale due to chiral symmetry breaking and the twist--4 
and 3 quark--gluon correlations governing the $1/Q^2$ corrections to 
the nucleon spin structure functions. In Sec.~\ref{sec:size}, we argue 
that, on general grounds, the flavor--nonsinglet twist--4 matrix element, 
$f_2^{u - d}$, has a sizable negative value of the order $\rho^{-2}$, 
due to the presence of sea quarks with virtualities $\sim \rho^{-2}$ 
in the nucleon wave function. The twist--3 matrix element, $d_2$, 
however, is not related to the scale $\rho^{-2}$. Our arguments 
provide additional insights into the origin of the instanton vacuum 
results, and suggest that they may be of more general nature. 
In Sec.~\ref{sec:fit}, we present the results of a next--to--leading order 
(NLO) QCD analysis of the world data for $g_1^p$ and $g_1^n$, including 
recent data from the Jefferson Lab Hall A \cite{Zheng:2003un} and 
COMPASS \cite{Ageev:2005gh} experiments, in which we
extract the flavor--nonsinglet twist--4 matrix element, $f_2^{u - d}$. 
The sign and order--of--magnitude are found to be in agreement with the
our qualitative arguments (as well as the instanton vacuum results). 
This analysis extends previous phenomenological estimates of the higher--twist
contribution to polarized deep--inelastic scattering by Leader et
al. \cite{Leader:2002ni,Leader:2004qv}. In contrast to other 
analyses \cite{Deur:2004ti} we perform the QCD fit to the data 
(including higher--twist corrections) for the
$x$--dependent structure function, computing the moments only at
the last stage, by integration of the fit.  Furthermore, we extract
directly the isovector (flavor--nonsinglet) twist--4 matrix element,
which is practically scheme--independent and thus provides a much cleaner 
probe of the non-perturbative quark--gluon correlations than the singlet
matrix element. Finally, in Sec.~\ref{sec:discussion}, 
we comment on the implications of the ``two--scale
picture'' of hadron structure for the $x$--dependence of the
higher--twist contribution and quark--hadron duality in the
spin structure functions.
\section{Higher--twist matrix elements and the size of the 
constituent quark}
\label{sec:size}
\begin{figure}[b]
\includegraphics[width=6cm]{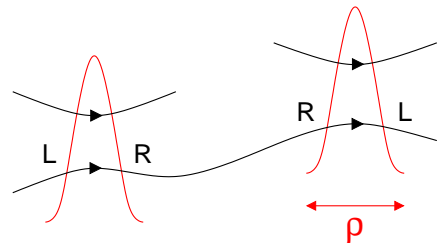}
\caption[*]{Origin of the short--distance scale, $\rho$, associated with
chiral symmetry breaking.}
\label{fig:propag}
\end{figure}
The spontaneous breaking of chiral symmetry implies the existence
of a short--distance scale in the QCD vacuum, significantly smaller 
than the typical hadronic size. It is determined by the characteristic
size, $\rho$, of the non-perturbative gluon field configurations which change
the quark chirality, see Fig.~\ref{fig:propag}. 
A gauge--invariant measure of this size is the ratio
of the dimension--5 ``mixed'' quark--gluon condensate to the usual 
dimension--3 quark condensate,
\begin{equation}
m_0^2 \;\; \equiv \;\; 
\frac{\langle\bar\psi \sigma^{\alpha\beta} g G_{\alpha\beta} \psi\rangle}
{\langle\bar\psi \psi\rangle} \\
\;\; = \;\; \frac{2 \, \langle\bar\psi \nabla^2 \psi\rangle}
{\langle\bar\psi \psi\rangle} .
\label{m_0^2}
\end{equation}
Here $\sigma^{\alpha\beta} \equiv (i/2)[\gamma^\alpha , \gamma^\beta ]$, and
$\nabla_\alpha \equiv \partial_\alpha - i g A_\alpha$ is the
covariant derivative. The two representations of the
dimension--5 operator are related by the identity 
\begin{equation}
g G_{\alpha\beta} \;\; = \;\;
i \left[ \nabla_\alpha , \nabla_\beta \right],
\label{identity}
\end{equation}
and the Heisenberg equations of motion for the quark fields.
Lattice simulations indicate that $m_0^2 \gtrsim 1\, \text{GeV}^2$ 
at a normalization point of $\mu \sim 1\, \text{GeV}$
\cite{Kremer:1987ve,Chiu:2003iw} (substantially larger values
were obtained in Ref.~\cite{Doi:2002wk}). A more precise 
interpretation of the ratio (\ref{m_0^2}) in terms of a size 
of field configurations becomes possible with specific assumptions
about the shape of these configurations in a given gauge.
In the instanton vacuum \cite{Diakonov:2002fq,Schafer:1996wv}, 
where the chirality--flipping field configurations are (anti--)
instantons in singular gauge, one has
\begin{equation}
m_0^2 \;\; = \;\; 4 \rho^{-2} ,
\end{equation}
at the scale $\mu \sim \rho^{-1}$ \cite{Polyakov:1996kh}.
The lattice results are consistent with an average instanton size of
$\rho \approx 0.3 \, \text{fm}$. We note that in the presence of 
more than one light quark flavor, the chirality flip due to the
instanton happens in a many--fermionic interaction ('t Hooft vertex),
whence the scale $\rho$ may also be interpreted as the range of
chiral--symmetry breaking quark--quark interactions in the QCD vacuum.

A simple heuristic argument suggests that the twist--4 matrix element, 
$f_2$, Eq.~(\ref{f2_def}), may be related to the short--distance scale,
$\rho$. Substituting in the twist--4 operator the gluon field by the 
commutator of covariant derivatives,
Eq.~(\ref{identity}), and making use of gamma matrix identities and the
Heisenberg equations of motion of the quark fields, one can convert
the twist--4 operator to the form \cite{Ehrnsperger:1993hh}
\begin{equation}
f_2 : \;\;\;\; \bar\psi \gamma^\beta \gamma_5 (-\nabla^2 ) \psi .
\label{axial_virtuality}
\end{equation}
In this form, it can be compared with the axial current operator,
which measures the quark contribution to the nucleon spin,
\begin{equation}
g_A : \;\;\;\; \bar\psi \gamma^\beta \gamma_5 \psi .
\label{axial}
\end{equation}
We see that the operator (\ref{axial_virtuality}) measures the correlation 
of the spin of the quarks with the square of their canonical momentum.
The existence of the short--distance scale, $\rho$, implies that, generally
speaking, nucleon correlation functions involve ``sea'' quarks with 
virtualities (four--momenta squared) up to the scale $\rho^{-2}$. If the 
spin of these quarks is correlated with the nucleon spin, one would
expect the flavor--nonsinglet twist--4 matrix element to be of the order
\begin{equation}
f_2^{u - d} \;\; \sim \;\; g_A \; \rho^{-2} .
\label{f_2_parametric}
\end{equation}
Here we limit ourselves to the flavor--nonsinglet ($u - d$) operators, 
which are not affected by the $U(1)_A$ anomaly (henceforth, $g_A$
denotes the nucleon's isovector axial coupling). Furthermore, if 
the flavor--nonsinglet sea quarks were polarized along
the direction of the nucleon spin, and if the quark canonical momentum 
squared on average takes positive values, as suggested by Eq.~(\ref{m_0^2}),
one would conclude that $f_2^{u - d} < 0$.

%
%
\begin{figure}[t]
\includegraphics[width=8.0cm]{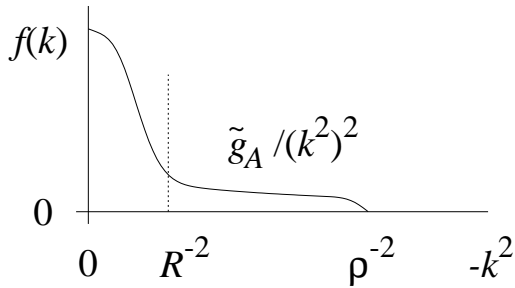}
\caption[*]{Distribution of quark virtualities, $k^2$, in the 
proton's isovector axial charge, as given by the effective
chiral theory, Eq.(\ref{ga_distribution}). The twist--4 matrix
element, $f_2^{u - d}$, is of the order of the integral of 
$k^2 \, f(k)$.}
\label{fig:gadist}
\end{figure}
The above argument supposes the existence of a dynamical mechanism
which polarizes the flavor--nonsinglet ``sea'' quarks in the nucleon.
In fact, such a mechanism can be found in the effective dynamics 
resulting from the spontaneous breaking of chiral symmetry. 
In the large--$N_c$ limit of QCD, the effective dynamics at distances
$\gtrsim \rho$ can be described by massive ``constituent''
quarks, coupled to a Goldstone pion field in a chirally invariant way
\cite{Diakonov:1984tw,Diakonov:1985eg},
\beq  
L_{\text{eff}} \;\; = \;\; \bar\psi (x) \left[ i\hat\partial \; - \; M 
e^{i \gamma_5 \pi^a (x) \tau^a} \right]
\psi(x) .  
\label{L_eff} 
\eeq  
The dynamical quark mass, and the coupling to the pion field,
are active for quark virtualities $|k^2| \lesssim \rho^{-2}$, whence 
$\rho \ll M^{-1}$ can be interpreted as the ``size'' of the constituent 
quark. The large--$N_c$ limit implies a semiclassical description
of the nucleon, in which the nucleon is characterized by a classical
pion field of size $R \sim M^{-1}$ \cite{Diakonov:1987ty}. 
Thus, the effective dynamics described by Eq.~(\ref{L_eff}) embodies 
the ``two--scale picture'' of the nucleon described in the 
introduction, Eq.~(\ref{hierarchy}).

%
%
\begin{figure}[t]
\includegraphics[width=4cm]{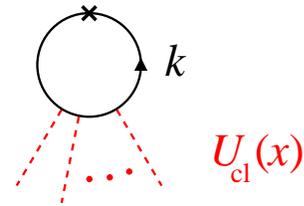}
\caption[*]{The ``sea'' quark contribution to the axial charge, 
giving rise to the large--virtuality tail ($-k^2 \sim \rho^{-2}$) 
in the distribution $f(k)$, Eqs.~(\ref{ga_power}) and (\ref{ga_grad}).}
\label{fig:sea}
\end{figure}
Consider now the matrix element of the flavor--non\-singlet axial current 
operator, \textit{i.e.}, the nucleon's isovector axial charge, in the
effective theory of constituent quarks coupled to a pion field, 
Eq.~(\ref{L_eff}). After appropriate projections on nucleon states with 
definite spin/flavor/momentum quantum numbers, the matrix element can be 
represented in the form
\beq
g_A \;\; = \;\; \int\frac{d^4 k}{(2\pi )^4} \; f(k),
\label{ga_distribution}
\eeq
where $k$ represents the four--momentum of the quark connected 
to the axial current operator. We are interested in the distribution
of quark virtualities, $k^2$, in this integral \footnote{Here $k$ denotes
the Minkowskian 4--momentum (all our formulas assume the Minkowskian metric).
The integration effectively extends over spacelike (Euclidean) 
momenta, for which $-k^2 = k_{\text{Euclid}}^2 > 0$.}. 
This distribution consists of two components,
shown schematically in Fig.~\ref{fig:gadist}.
The bulk of the nucleon's axial charge is carried by valence and sea 
quarks with virtualities of the order of the size of the nucleon, 
$-k^2 \sim R^{-2}$. In addition, however, there is a contribution from
``sea'' quarks interacting with the classical pion field,
which have virtualities extending up to $-k^2 \sim \rho^{-2}$. 
Assuming that $\rho \ll R$, the power behavior and the coefficient
of the this large--virtuality ``tail'' are completely determined by 
general features of the chiral dynamics. They follow from the
leading term in the long--wavelength expansion (gradient expansion)
of the quark loop in the background pion field (see Fig.~\ref{fig:sea}):
\beq
f(k) \;\; \sim \;\; \frac{\tilde g_A}{(k^2)^2} 
\hspace{4em} (-k^2 \sim \rho^{-2}),
\label{ga_power}
\eeq
where the coefficient is given by
\beq
\tilde g_A \;\; \equiv \;\; \frac{4 N_c M^2}{9} \int d^3 x \;   
{\rm tr}\,\left[ -i \tau^a U_{\text{cl}}^\dagger (\bm{x}) 
\partial_a U_{\text{cl}} (\bm{x}) \right] .  
\label{ga_grad}  
\eeq   
Here $U_{\text{cl}} (\bm{x}) \equiv 
\exp \left[ i \pi^a_{\text{cl}} (\bm{x}) \tau^a \right]$ 
denotes the static classical pion field in the
nucleon rest frame. In fact, the expression (\ref{ga_grad}) coincides
with the leading--order gradient expansion of the axial charge
induced by the classical pion field (as used \textit{e.g.} for
calculating $g_A$ in the Skyrme model).
The power behavior (\ref{ga_power}) implies that the integral
(\ref{ga_distribution}) depends only logarithmically on the 
short--distance scale $\rho$:
\begin{equation}
g_A \;\; \sim \;\; \log \frac{\rho}{R} .
\label{g_A_integral}
\end{equation}
Thus, $g_A$ is dominated by quark virtualities of the order 
$R^{-2} \ll \rho^{-2}$, and the presence of the large--virtuality
tail in the distribution of Fig.~\ref{fig:gadist} is of minor importance.

The presence of the large--virtuality tail in the axial charge distribution
becomes crucial, however, for the twist--4 matrix element, $f_2^{u - d}$.
When passing from QCD to the effective theory of constituent quarks,
QCD operators must be ``translated'' to operators in the effective theory. 
It is natural to assume that in the case of the twist--4 QCD 
operator (\ref{axial_virtuality}) the translation is given by
\beq
\bar\psi \gamma^\beta \gamma_5 (-\nabla^2 ) \psi \, |_{\text{QCD}}
\;\; \rightarrow \;\;
C \, \bar\psi \gamma^\beta \gamma_5 (-\partial^2 ) \psi 
\, |_{\text{eff}} ,
\label{matching}
\eeq
where $C > 0$ is a coefficient of order unity. This was explicitly
demonstrated in the instanton vacuum model, where the operator matching
follows from the integration over instanton--type gauge field 
configurations \cite{Balla:1997hf}. By analogy with the
axial current, the matrix element of the constituent quark
operator (\ref{matching}) is now given by the integral of $k^2$
times the momentum distribution of $g_A$, Eq.~(\ref{ga_distribution})).
This factor suppresses contributions from virtualities $-k^2 \sim R^{-2}$,
and enhances the contributions from the large--virtuality tail.
As a result, the twist--4 matrix element is parametrically of the order
\begin{equation}
f_2^{u - d} \;\; \sim \;\; \rho^{-2} ,
\label{f_2_sign}
\end{equation}
with the coefficient proportional to $\tilde g_A$, in agreement
with our estimate (\ref{f_2_parametric}). Furthermore, since the 
quark virtualities in the ``tail'' of the distribution 
are spacelike, $k^2 < 0$, we conclude that $f_2^{u - d} < 0$.

The integral determining $f_2^{u - d}$ in the effective chiral theory
contains a would--be quadratic divergence, which is cut off at the
scale $\rho^{-2}$. While the parametric order and sign of $f_2^{u - d}$
follow from general features of the chiral dynamics,
the numerical value is very sensitive to the precise way in which
the UV cutoff (\textit{i.e.}, the finite size of the constituent quark)
is implemented. The instanton vacuum model, which implies a definite
form of the UV cutoff, resulting from the fermionic zero modes of the
instantons \cite{Diakonov:1985eg}, and which also allows one to uniquely 
determine the matching coefficient for the effective twist--4 
operator, Eq.~(\ref{matching}) \cite{Balla:1997hf}, 
gives (with $\rho = 0.3\, \text{fm}$) \cite{Lee:2001ug}
\beq  
f_2^{u - d} \;\; \approx \;\; -0.5 \, g_A \, \rho^{-2} \;\; = \;\; 
-0.22 \, \text{GeV}^2 .  
\label{f2_instanton}
\eeq  
However, in view of the principal uncertainties in the modeling
of the dynamics of quark field modes of virtualities $\sim \rho^{-2}$
this result should be viewed as a rough estimate. 

On general grounds, one expects that in QCD 
with cutoff regularization the matrix elements of twist--4 operators are 
proportional to the square of the cutoff \cite{Martinelli:1996pk}; this is 
realized in our approach if the cutoff is identified with the scale 
$\rho^{-1}$. We do not consider here the logarithmic scale dependence of 
the twist--4 matrix element which results from the standard composite operator 
renormalization \cite{Shuryak:1981pi}; this dependence is much weaker than 
the principal uncertainty in our estimates of the twist--4 matrix element.

It is interesting that the estimates of $f_2^{u - d}$ obtained in
QCD sum rule calculations \cite{Balitsky:1989jb,Stein:1995si}
agree in sign and order--of--magnitude with our estimate, and with the
instanton vacuum result, see Table~\ref{table} (for a critical discussion
of these calculations, see Ref.~\cite{Ioffe:1996ey}). Our estimate
disagrees with the bag model \cite{Ji:1993sv}, which gives a positive 
result for $f_2^{u - d}$ (note the different convention for the sign
of $f_2$ in that paper). This model, however, does not respect the 
QCD equations of motion, and therefore cannot claim to give a realistic 
description of quark--gluon correlations in the nucleon.
%
%
\begin{table}[t]
\begin{tabular}{l|l}
                     & \,\,\, $f_2^{u - d} \, [{\rm GeV}^2]$ \\ \hline
Instantons \cite{Balla:1997hf,Lee:2001ug}    & \,\,\, $-0.22$  
\\ \hline
QCD sum rules (Balitsky et al.) 
\cite{Balitsky:1989jb}                       & \,\,\, $-0.16 \pm 0.04$ 
\\ \hline
QCD sum rules (Stein et al.) \cite{Stein:1995si}        
                                             & \,\,\, $-0.06 \pm 0.02$ 
\\ \hline
Bag model \cite{Ji:1993sv}                   & \,\,\, +0.09   \\ \hline
\end{tabular}
\caption[]{Comparison of theoretical estimates of the flavor--nonsinglet 
twist--4 proton matrix element, $f_2^{u - d}$.}
\label{table}
\end{table}

When applying the same reasoning as above to the twist--3 matrix element,
$d_2$, we find that after the substitution (\ref{identity})
the quark--gluon operator does not produce a contracted covariant
derivative. In this operator, all derivatives are ``kinematic'', {\it i.e.},
they are needed to support the spin of the matrix element.
This operator does not probe the virtuality of the quarks in the 
nucleon, and its matrix element does not receive essential contributions
from quark virtualities of the order $\rho^{-2}$. Thus, the correlations
probed by the twist--3 matrix element are of essentially different nature
as those probed by the twist--4 one. Beyond this qualitative difference,
we see no simple way to estimate the twist--3 matrix element more
accurately on grounds of general features of the effective dynamics
alone (see also Ref.~\cite{Wandzura:1977qf}). In the instanton vacuum, 
an additional suppression of $d_2$ results from the fact that the 
coefficient of the corresponding effective operator in the effective 
chiral theory is parametrically small in the instanton packing fraction
(\textit{i.e.}, $d_2$ is zero in the single instanton approximation).
This suppression appears to be due to the $O(4)$ invariance
(in the Euclidean metric) of the instanton field \cite{Balla:1997hf}.
In a sense, the twist--3 operator is a much more subtle probe 
of non-perturbative quark--gluon correlations in the QCD vacuum
than the twist--4 operator, whose matrix element can be estimated
on general grounds.
\section{Twist--4 matrix element from a QCD fit
to polarized DIS data}
\label{sec:fit}
It is interesting to see to which extent our qualitative conclusions
about the higher--twist matrix elements are supported by the 
experimental data. To this end, we attempt to extract the
flavor--nonsinglet twist--4 matrix element, $f_2^{u - d}$, 
from the power corrections to the spin structure functions $g_1^p$
and $g_1^n$. To accuracy $1/Q^2$, the tree--level QCD expansion 
for the first moment of $g_1^p - g_1^n$ is given 
by \cite{Shuryak:1981pi,Ji:1993sv,Ehrnsperger:1993hh}
\be
\int_0^1 \! dx \, g_1^{p - n} (x, Q^2 ) &=&
\frac{a_0^{u - d}}{6} + \frac{M_N^2}{27 Q^2} \, a_2^{u-d} 
\nonumber \\
&+& \frac{4}{27 Q^2} \left( M_N^2 \, d_2^{u-d} + f_2^{u - d} \right) 
\;\;\;\;
\label{g_1_QCD}
\ee
(when QCD radiative corrections are included, the coefficients 
acquire a logarithmic $Q^2$--dependence).
The first term is the leading--twist (LT) contribution, proportional 
to the matrix element of the flavor--nonsinglet twist--2 spin--1 operator, 
$a_0^{u - d}$, with $a_0^{u - d} \equiv g_A$ (Bjorken sum rule).
The second term represents the target mass corrections (TMC),
proportional to the spin--3 twist--2 matrix element, $a_2^{u - d}$.
The third term is the dynamical higher twist (HT) contribution,
involving the twist--3 and 4 matrix elements (\ref{d2_def}) and 
(\ref{f2_def}). The twist--3 matrix element, $d_2$, has been 
extracted from independent measurements of the third moment of 
the spin structure function $g_2$ (with the
Wandzura--Wilczek contribution subtracted). The SLAC E155X
experiment \cite{Anthony:2002hy} and the recent Jefferson Lab Hall A
analysis \cite{Zheng:2004ce} report values of $d_2^{p, n} \lesssim 10^{-2}$,
in good agreement with the instanton vacuum prediction \cite{Balla:1997hf}. 
With these values, the contribution of the $M_N^2 d_2^{u - d}$ term
to the $1/Q^2$ corrections in (\ref{g_1_QCD}) is more than an order of 
magnitude smaller than the instanton estimate for $f_2^{u - d}$, 
Eq.~(\ref{f2_instanton}). This supports the qualitative conclusion from 
our two--scale picture, that the dominant power corrections are those
associated with the short--distance scale, $\rho$. We shall thus
neglect the $M_N^2 d_2^{u - d}$ term compared to $f_2^{u - d}$ in 
Eq.~(\ref{g_1_QCD}), and ascribe the phenomenological power correction
entirely to the twist--4 matrix element. This theoretical simplification
will be justified \textit{a posteriori} by the fact that the numerical 
value of $f_2^{u - d}$ extracted in this way is indeed much larger
than the measured $M_N^2 d_2^{u - d}$.

\begin{figure}[t]
\includegraphics[width=8.6cm]{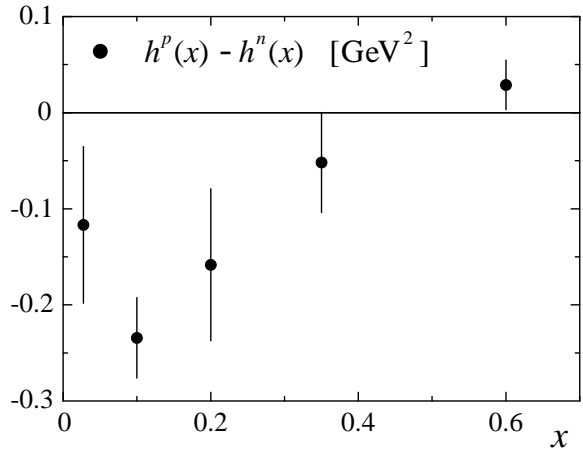}
\caption[*]{The difference of higher twist corrections to the 
proton and neutron spin structure functions, $h^{p}(x) -  h^{n}(x)$, 
Eq.~(\ref{g_1_fit}), as extracted from our NLO QCD fit to the
world data (see references in Ref.~\cite{Leader:2005kw}), 
including recent $g_1^n$ data from the Jefferson Lab Hall A 
experiment \cite{Zheng:2003un}, as well as deuteron data 
from COMPASS \cite{Ageev:2005gh}.}
\label{fig:fit}
\end{figure}

The dynamical higher--twist contribution to the $x$--dependent
structure functions, $g_1^p (x, Q^2)$ and $g_1^n (x, Q^2)$, 
has been extracted from NLO QCD fits to the world data 
(see references in Ref.~\cite{Leader:2005kw}),
including the new $g_1^n$ data from the Jefferson Lab Hall A
experiment \cite{Zheng:2003un}, as well as the deuteron data from 
COMPASS \cite{Ageev:2005gh}. These fits are based on the ansatz
\begin{equation}
g_1^{p,n} (x, Q^2) \;\; = \;\; g_1^{p,n} (x,
Q^2)_{\mbox{{\scriptsize LT + TMC}}} \; + \; \frac{h^{p,n}
(x)}{Q^2} , 
\label{g_1_fit}
\end{equation}
where the leading--twist contribution (including target mass
corrections) is given by the Leader--Stamenov--Sidorov
parametrization of the polarized parton densities \cite{Leader:2005kw}, 
and $h^{p(n)}$ parametrizes the dynamical higher--twist corrections.
In order to extract directly the flavor--nonsinglet higher--twist
correction, we have modified the fit procedure of 
Refs.~\cite{Leader:2002ni,Leader:2005kw} and parametrized not $h^{p(n)}(x)$
individually, but their difference and sum. 
The kinematic cuts applied in the new fit are the same as in 
the fit of Ref.~\cite{Leader:2005kw}, $Q^2 \geq 1 \,\text{GeV}^2$ 
and $W^2 \geq 4\, \text{GeV}^2$. The results for the difference, 
$h^{p}(x) - h^{n}(x)$, obtained in this way
is shown in Fig.~\ref{fig:fit}.
%
%

Integrating the higher--twist contribution over the $x$--range
covered in our fit we get
\begin{equation}
\int_{0.017}^{0.75} dx \; [h^{p} (x) -  h^{n} (x)]  \;=\; -0.046
\pm 0.016 \; {\rm GeV}^2 \label{NS_int} . \\
\label{x_integral_partial}
\end{equation}
If we neglected the contribution from the large--$x$ region 
(we shall argue in Section~\ref{sec:discussion} that this
is consistent with our two--scale picture of the structure of the nucleon),
as well as the small--$x$ region, and regarded the integral 
(\ref{x_integral_partial}) as an estimate of the first moment, 
we would obtain [\textit{cf.}\ Eq.~(\ref{g_1_QCD})]
\beq
f_2^{u - d} \;\; = \;\; - 0.31 \pm 0.11 \; {\rm GeV}^2 .
\label{f2_fit_simple}
\eeq
For a rough data--based estimate of the contribution of the large--$x$ 
region, we consider the integrals up to $x = 1$ computed with 
a linear extrapolation of the fit into this region, based on the two 
rightmost data points in Fig.~\ref{fig:fit}, and a constant extrapolation,
based on the rightmost data point only (we neglect the small--$x$ 
contribution). Taking the average of the two integrals as an estimate 
of the central value, and half the difference as an estimate of the error, 
we get 
\begin{equation}
\int_{0}^{1} dx \; [h^{p} (x) -  h^{n} (x)]  \;=\; -0.028
\pm 0.019 \; {\rm GeV}^2 ,
\label{NS_int_tot} \\
\end{equation}
corresponding to 
\begin{equation}
f_2^{u - d} \;\; = \;\; -0.20 \pm 0.14 \; {\rm GeV}^2 .
\label{f_u_minus_d_fit}
\end{equation}
Both estimates (\ref{f2_fit_simple}) and (\ref{f_u_minus_d_fit})
agree in sign and in order--of--magnitude with our qualitative 
prediction, Eq.~(\ref{f_2_parametric}), and the result obtained 
in the instanton vacuum, Eq.~(\ref{f2_instanton}).

In the QCD fit, the leading--twist parton densities and coefficient 
functions were taken in the $\overline{MS}$ scheme. The studies of 
Ref.~\cite{Leader:2002ni} found that the higher--twist corrections 
to $g_1$ in different factorization schemes (in particular, in the 
$\overline{MS}$ and JET schemes) coincide within errors. Note also that
Eq.~(\ref{f2_fit_simple}) agrees well with our previous estimate
based on the JET scheme \cite{Sidorov:2004sg}. In that estimate,
also preliminary HERMES data on the deuteron structure function
were taken into account \cite{Weiskopf:2002yw}. The recent analysis
of Ref.~\cite{Leader:2005ci} shows that inclusion of the new 
HERMES proton and deuteron data \cite{Airapetian:2004zf} 
does not substantially change the higher--twist contribution,
whence we have not explicitly included these data in our 
present analysis.

Our result for $f_2^{u - d}$ agrees well with that obtained by
Deur {\it et al.} \cite{Deur:2004ti} in a recent analysis of
power corrections to the Bjorken sum rule
(their $f_2^{p - n} \equiv \frac{1}{3} f_2^{u - d}$ in our conventions).
Our method differs from
that of Ref.~\cite{Deur:2004ti} in that we perform the QCD fit to
the data (including higher--twist corrections) for the
$x$--dependent structure function, computing the moments only at
the last stage, by integration of the fit. Nevertheless, the
results for the higher--twist contribution from both approaches
are comparable, which is very encouraging. Our result disagrees 
in sign with that obtained in an earlier analysis of the 
$Q^2$--dependence of the Bjorken sum rule \cite{Kao:2003jd},
which combined an empirical parametrization of the electroproduction
cross section in the resonance region with the QCD parametrization
in the DIS region, see also Refs.~\cite{Ji:1997gs,Edelmann:1999yp}.
\section{Discussion and outlook}
\label{sec:discussion}
In our discussions so far we considered the twist--4 correction
to the first moment of $g_1(x)$, which is related to the matrix element 
of the local twist--4, spin--1 operator, $f_2$. Much more information
is contained in the $x$--dependence of the higher--twist contribution.
A detailed study of the $x$--dependence of the twist--4 contribution
on the basis of chiral dynamics and the instanton vacuum is beyond the
scope of the present paper. Here we would like to offer only some general
comments on this problem.

The idea of a two--scale picture of the nucleon outlined in 
Section~\ref{sec:introduction}, \textit{cf.}\ Eq.~(\ref{hierarchy})
can also be expressed in a partonic language. In this formulation,
constituent quarks/antiquarks appear as correlations in the transverse
spatial distribution of quarks/antiquarks and gluons, with a transverse
size, $\rho$, significantly smaller than the transverse size of the
fast--moving nucleon, $R$. (This formulation is in fact used to discuss 
the effect of the constituent quark structure of the nucleon on
high--energy $pp$ scattering with multiple hard processes 
\cite{Frankfurt:2004kn}.)  This picture applies to average values
of the quark/antiquark longitudinal momentum fraction, $x \lesssim 0.5$, for 
which the transverse size of configurations in the nucleon wave 
function is of the order of the typical hadronic size. In this
formulation, the twist--4 correction to the spin structure function
can be related to the average transverse momentum squared of 
the polarized quarks/antiquarks in the nucleon \cite{Ellis:1982wd,Ji:1993ey}.
Extending the reasoning of Section~\ref{sec:size}, one would argue
that due to chiral dynamics the flavor--nonsinglet polarized ``sea'' 
quark distribution involves transverse momenta squared of the order 
$k_\perp^2 \sim \rho^{-2}$. This suggests an interesting connection
between the flavor--nonsinglet twist--4 corrections to $g_1$ and
the large flavor asymmetry of the twist--2 sea quark distribution, 
$\Delta\bar u(x) - \Delta \bar d(x)$ predicted by chiral 
dynamics in the large--$N_c$ limit \cite{Diakonov:1996sr}. 
The result of our QCD fit, Fig.~\ref{fig:fit}, indicates that 
the twist--4 correction is indeed localized at relatively small 
values of $x \sim 0.1$, supporting the connection with the sea 
quark distribution. A similar connection between the flavor asymmetry
and the presence of large transverse momenta in the sea quark
distributions was noted in Ref.~\cite{Dorokhov:1993fc}, where
instanton--induced sea quark components in the nucleon wave function 
were considered in a phenomenological model with no reference to the 
large--$N_c$ limit.

The existence of the short--distance scale due to chiral symmetry
breaking also has some interesting qualitative implications for 
quark--hadron duality in polarized DIS.
The two--scale picture of the structure of the nucleon, \textit{cf.}\
Eq.~(\ref{hierarchy}), implies a parametric 
classification of the hadronic excitation spectrum of the nucleon. 
Nucleon resonances such as the $\Delta$ are excitations with energies 
(in the CM frame) of the order $E \sim R^{-1}$. They correspond to changes 
of the state of motion of the constituent quarks over distances $\sim R$, 
which do not affect the internal structure of the constituent quark at 
distances $\sim \rho$. Excitations of energy $E \sim \rho^{-1}$ belong 
to the non-resonant hadronic continuum. Switching to the quark language,
our arguments of Section~\ref{sec:size} show that the twist--2 quark
distribution (here, the axial coupling, $g_A$) arises mainly from 
field configurations with energies/momenta of the order $R^{-1}$, 
while the twist--4  quark--gluon correlations are dominated by 
energies/momenta of the order $\rho^{-1}$. Comparing the hadronic
and the quark description, we conclude that quark--hadron duality 
should ``work'' for the twist--2 part of the structure function 
(\textit{i.e.}, the $Q^2$--independent part) already when summing 
over hadronic excitations with energies $E \sim R^{-1}$, but 
for the twist--2 plus twist--4 part of the structure function
(\textit{i.e.}, to accuracy $1/Q^2$) only when summing over hadronic 
excitations with energies $E \sim \rho^{-1}$. In practice, this means
that quark--hadron duality in the structure function to accuracy 
$1/Q^2$ may require integration over a significantly larger duality 
interval than duality to accuracy $(Q^2)^0$. This needs to be taken
into account when trying to extract higher--twist matrix elements
from resonance--based parametrizations of the structure functions.
To summarize, the two--scale picture makes a clear parametric distinction 
between resonance and higher--twist contributions to the structure function. 
This qualitative prediction is supported by the fact that the 
phenomenological twist--4 contribution to $g_1$ (see Fig.~\ref{fig:fit})
seems to be dominated by small values of $x$, below the resonance region.

The two--scale picture described here is close in spirit to the 
Ioffe--Burkert parametrization of the $Q^2$ dependence of the first moment
of $g_1^p$ \cite{Burkert:1992tg}, in which the contribution
from the $\Delta$ resonance is separated from the continuum, and
the leading power corrections are associated with the
continuum contribution. The characteristic mass scale governing 
the power corrections in this parametrization, $\mu^2 = M_\rho^2$,
is numerically close to value associated with the constituent quark 
size, $\rho^{-2} \sim (0.3 \, {\rm fm})^{-2} = (600 \, {\rm MeV})^2$.
Also, the analogous parametrization for the Bjorken sum rule 
$(p - n)$ gives negative sign of the twist--4 correction,
in agreement with our qualitative prediction. However, the numerical value
of the twist--4 correction obtained from the Ioffe--Burkert parametrization 
is substantially larger than the instanton vacuum estimate, corresponding 
to $f_2^{u - d} \approx - 2.3 \, g_A \, \mu^2 = 1.7 \; \text{GeV}^2$. 

More generally, the two--scale picture of hadron structure allows one 
to draw some conclusions about global properties of the transition 
from high to low $Q^2$ in the nucleon spin structure functions
({\it i.e.}, going beyond the leading $1/Q^2$ corrections).
Since the characteristic mass scale for the power corrections
is set by the size of the constituent quark,
one should expect the twist expansion
to break down at momenta of the order $Q^2 \sim \rho^{-2}$.
For the extraction of the leading ($1/Q^2$--) corrections from
QCD fits to the data this implies that one should restrict oneself
to the range $Q^2 \gg \rho^{-2}$, where the leading term in the
series dominates (in our fit presented in Section~\ref{sec:fit},
$Q^2 > 1\, \text{GeV}^2$).

The arguments presented in this paper can also be extended
to higher--twist corrections to unpolarized deep--inelastic scattering.
A new feature in the unpolarized case is the appearance of twist--4 
operators measuring quark--quark correlations (``four--quark operators''), 
in addition to quark--gluon correlations of the type encountered 
in the polarized structure functions. Within our two--scale picture,
the quark--quark correlation matrix elements are of the order $R^{-2}$,
and thus parametrically suppressed compared to the quark--gluon ones,
which are of the order $\rho^{-2}$. This qualitative conclusion seems
to be in agreement with a joint analysis of the twist--4 corrections
to $F_2$ and $F_L$, see Ref.~\cite{Dressler:1999zi} for details.
The role of the size of the constituent quark in power corrections
to unpolarized structure function moments was also discussed in
a different approach in Ref.~\cite{Petronzio:2003bw}.

To summarize, we have argued that the leading power corrections
to the nucleon spin structure functions are governed by the 
short--distance scale due to the spontaneous breaking of chiral
symmetry --- the size of the constituent quark. The qualitative
statements following from this assumption are supported by
the result of a QCD fit to the present polarized DIS data.
The arguments presented here may eventually serve as the basis for
an ``interpolating'' model of the nucleon spin structure functions,
connecting the scaling region at large $Q^2$ with the 
photoproduction point.

We thank J.--P.~Chen, W.~Melnitchouk, D.~Stamenov, and 
M.~Vanderhaeghen for useful discussions. 
This work is supported by U.S.\ Department of Energy 
contract DE-AC05-84150, under which the Southeastern Universities
Research Association (SURA) operates the Thomas Jefferson National
Accelerator Facility. A.~V.~S.\ acknowledges financial support by
RFBR (Grants 05-02-17748, 05-01-00992, 06-02-16215, and 06-02-81032) 
and the Heisenberg--Landau program.
\end{document}